\documentclass[aps,prb,reprint,superscriptaddress,longbibliography]{revtex4-2}

\usepackage{amsmath}

\usepackage{amssymb}

\usepackage{graphicx}

\usepackage{bm}

\usepackage{hyperref}

\begin{document}

\title{Multi-image Overlap Stitching and Automatic Image Construction for Coherent X-ray Imaging}

\author{Starr Boney}

\affiliation{Department of Physics and Astronomy, University of Wisconsin--Milwaukee, Milwaukee, Wisconsin 53211, USA}

\author{Umeshika Dissanayaka}

\affiliation{Department of Physics and Astronomy, University of Wisconsin--Milwaukee, Milwaukee, Wisconsin 53211, USA}

\author{Lillian Rutowski}

\affiliation{Department of Physics and Astronomy, University of Wisconsin--Milwaukee, Milwaukee, Wisconsin 53211, USA}

\author{Aaron George}

\affiliation{Department of Physics and Astronomy, University of Wisconsin--Milwaukee, Milwaukee, Wisconsin 53211, USA}

\author{Min Gyu Kim}

\email{mgkim@uwm.edu}

\affiliation{Department of Physics and Astronomy, University of Wisconsin--Milwaukee, Milwaukee, Wisconsin 53211, USA}

\date{\today}

\begin{abstract}

Direct-space and real-time coherent X-ray imaging (direct-CXI) enables visualization of magnetic-domain structures and dynamics over length scales exceeding the field of view of a single image. However, large-area measurements typically require raster scanning, producing hundreds of partially overlapping images that must be accurately aligned and combined before quantitative analysis can be performed. Here, we present Multi-image Overlap Stitching and Automatic Image Construction for coherent X-ray imaging (MOSAICX), an automated stitching workflow for large-area direct-space coherent X-ray imaging. The workflow consists of image centering, masking, trimming, binarization, hierarchical stitching, and post-processing. To demonstrate the method, we apply it to a dataset comprising 434 direct-CXI images of the antiferromagnetic topological insulator MnBi$_2$Te$_4$. The images are first combined into column reconstructions and subsequently stitched into a single large-area composite image. The resulting reconstruction reveals the complete magnetic-domain and domain-wall landscape over the scanned region while suppressing imaging artifacts and detector defects. The presented workflow provides an efficient approach for processing large direct-CXI datasets and enables visualization and analysis of magnetic-domain structures beyond the field of view of individual measurements.

\end{abstract}

\maketitle

\section{Introduction}

Antiferromagnetic (AFM) materials have emerged as promising candidates for spintronic applications owing to their potential for energy-efficient operation, ultrafast spin dynamics, and absence of stray magnetic fields.\cite{1,2,3,4,5,6,10,15,16} These advantages have stimulated extensive efforts to understand and control AFM order. Despite challenges such as low N\'eel temperatures and limited material availability, antiferromagnets exhibit a rich variety of magnetic phenomena that continue to attract considerable interest.

Magnetic domains and domain walls play a central role in determining the electrical, magnetic, and topological properties of AFM materials. Their morphology and dynamics can strongly influence material responses to external stimuli, making direct imaging of magnetic domains essential for both fundamental studies and emerging spintronic functionalities.\cite{1,15,1-1,10,16-0,16} This requires imaging techniques capable of resolving magnetic textures while maintaining a sufficiently large field of view to capture their collective behavior.

Existing AFM domain imaging techniques involve tradeoffs between spatial resolution, imaging speed, field of view, and material compatibility.\cite{10,16} Among recently developed approaches, direct-space and real-time coherent X-ray imaging (direct-CXI) provides a unique combination of large-area coverage and high temporal resolution, enabling direct observation of domain-wall motion and domain evolution in real time.\cite{7,12,13,14} Although its spatial resolution is lower than that of scanning probe techniques, direct-CXI is particularly well suited for investigating mesoscale magnetic textures and their dynamics.

Direct-CXI measurements are often performed by raster scanning the sample surface and acquiring a sequence of partially overlapping images over a large area. This acquisition strategy resembles that used in coherent X-ray ptychography;\cite{11-0,11-1,11-2,11-3,11} however, direct-CXI records images directly in real space and therefore does not require iterative reciprocal-space phase reconstruction.\cite{7,12,13,14} The resulting data provide rapid access to magnetic-domain configurations and dynamics over length scales exceeding the field of view of a single image. Nevertheless, a complete measurement may consist of hundreds of partially overlapping images that must be accurately aligned and combined into a composite image before quantitative analysis can be performed. The large number of images, together with imaging artifacts such as intensity gradients, detector defects, and optical distortions, makes manual stitching both time consuming and susceptible to cumulative alignment errors.

\begin{figure*}[t!]

\includegraphics[width=0.8\textwidth]{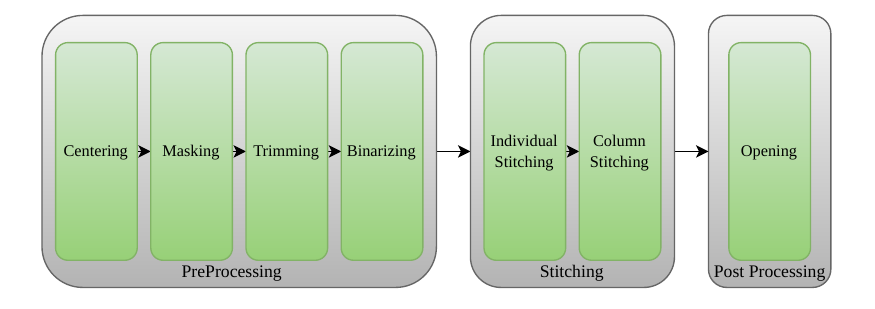}

\caption{Layout of the algorithm, organized into the three main stages and the processing steps within each stage.}

\label{fig1}

\end{figure*}

In this work, we develop an automated image-stitching procedure, Multi-image Overlap Stitching and Automatic Image Construction for coherent X-ray imaging (MOSAICX), for large direct-CXI datasets acquired using a Fresnel zone plate optical system. As a demonstration, we apply the method to direct-CXI measurements of the antiferromagnetic topological insulator MnBi$_2$Te$_4$, where magnetic domain walls appear as dark lines in the images. The dataset consists of 434 individual images, each containing $800 \times 800$ pixels, acquired in a raster pattern across the sample surface. To enable automated processing, the images are first preprocessed through centering, masking, trimming, and binarization procedures designed to remove imaging artifacts and enhance domain-wall contrast. The preprocessed images are then stitched in a hierarchical manner: individual images are first combined into column datasets, and the resulting columns are subsequently stitched to form a large-scale composite image. A final post-processing step is applied to remove residual stitching artifacts and improve image quality. Through MOSAICX, hundreds of individual direct-CXI images are transformed into a single large-area composite that reveals the complete magnetic-domain and domain-wall landscape inaccessible within the field of view of any individual image.

\section{Methods}

\subsection{Direct-CXI images}

Direct-CXI images were taken at the 23-ID-1 Coherent Soft X-ray Scattering (CSX) beamline of the National Synchrotron Light Source II at Brookhaven National Laboratory. This technique is phase sensitive and exploits destructive interference between neighboring antiferromagnetic (AFM) domains at their boundaries.\cite{7,12,13,14} Under magnetic Bragg diffraction conditions, ordered AFM domains produce strong scattering intensity, whereas interference between domains of opposite phase generates dark lines corresponding to domain walls.\cite{7,12,13,14} The incident X-ray energy was tuned to the Mn $L_3$ absorption edge ($E=640$ eV), and images were collected at the AFM Bragg peak $(0,0,1.5)$.

A single crystal of MnBi$_2$Te$_4$, grown at the University of Wisconsin--Milwaukee, with its crystallographic $c$ axis perpendicular to the sample surface was mounted on the cold finger of a liquid-helium flow cryostat installed on a $z$-axis diffractometer within the TARDIS ultra-high-vacuum instrument. Measurements were performed between 15 and 26 K. Snapshot images were acquired with exposure times ranging from 0.02 to 0.5 s. A Fresnel zone plate (FZP) was used as the imaging optic, and different magnifications were obtained by varying the sample-to-FZP distance.

The sample surface was measured using a raster-scanning procedure in which partially overlapping images were acquired at successive sample positions to image areas larger than the field of view of a single direct-CXI image. The dataset analyzed in this work consists of 434 images, each containing $800 \times 800$ pixels, arranged into 14 scan lines with 31 images per scan line. The corresponding image scale was calibrated by translating the sample relative to the X-ray optics using a high-precision SmarAct XYZ nanopositioning stage and measuring the resulting displacement of image features. This calibration provides a direct conversion between detector pixels and real-space distances on the sample surface.

\subsection{Image processing}

The MOSAICX workflow consists of three stages: preprocessing, stitching, and post-processing, as illustrated in Fig.~\ref{fig1}. During preprocessing, images are centered, masked, trimmed, and binarized to remove experimental artifacts and enhance domain-wall contrast. The preprocessed images are then stitched using a hierarchical procedure. Images acquired within each scan line are first combined to form column images, after which the column images are stitched together to generate a single large-area composite image. Finally, a post-processing procedure is applied to remove residual detector artifacts and improve image quality.

All image-processing steps were implemented using custom software\cite{Starr} developed for direct-CXI datasets. The workflow is designed to efficiently process large raster-scanned datasets and generate composite images suitable for visualization and quantitative analysis of magnetic-domain structures.

\section{Results}

\subsection{Overview of the stitching workflow}

Figure~\ref{fig1} summarizes the workflow used to construct large-area composite images from direct-CXI measurements of MnBi$_2$Te$_4$. The procedure consists of three stages: preprocessing, stitching, and post-processing. During preprocessing, experimental artifacts are removed and domain-wall contrast is enhanced. The processed images are then stitched using a hierarchical reconstruction procedure in which images acquired within the same scan line are first combined into column images, followed by stitching of the column images into a single large-area composite. Finally, post-processing is applied to remove residual detector artifacts and improve image quality.

\subsection{Preprocessing}

The first stage of the workflow consists of image centering, masking, trimming, and binarization. These procedures are applied independently to each scan line prior to stitching.

\subsubsection{Centering}

\begin{figure}[t]

\includegraphics[width=.7\columnwidth]{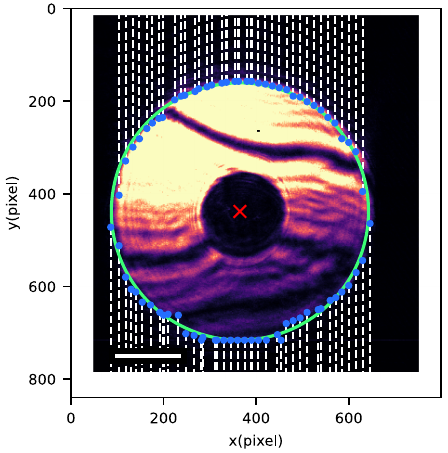}

\caption{Centering procedure. The virtual lines are indicated by the dotted lines. The detected edge points are indicated by the blue dots. The determined center is indicated by the red cross. The white scale bar indicates 5~$\mu$m.}

\label{fig2}

\end{figure}

The projected position of the direct-CXI image can vary from measurement to measurement because of surface roughness and small changes in the diffraction geometry. Consequently, all images must first be recentered before stitching can be performed.

Figure~\ref{fig2} illustrates the centering procedure. Multiple virtual lines (dashed lines in Fig.~\ref{fig2}) are projected from the image boundaries toward the donut-shaped intensity profile generated by the Fresnel zone plate optics. The positions (blue dots in Fig.~\ref{fig2}) where the intensity first deviates from the background are identified and used to define the boundary of the donut-shaped feature. These boundary points are subsequently fitted to a circle (green circle in Fig.~\ref{fig2}), and the center of the fitted circle is used as the image center (red cross in Fig.~\ref{fig2}). Each image is then translated so that all donut-shaped features are aligned to the same pixel coordinates.

Although this procedure significantly improves alignment, small variations in image intensity can cause slight differences in the detected boundary positions, leading to residual alignment errors that are addressed during the subsequent stitching procedure.

\subsubsection{Masking}

\begin{figure}[t]

\includegraphics[width=1\columnwidth]{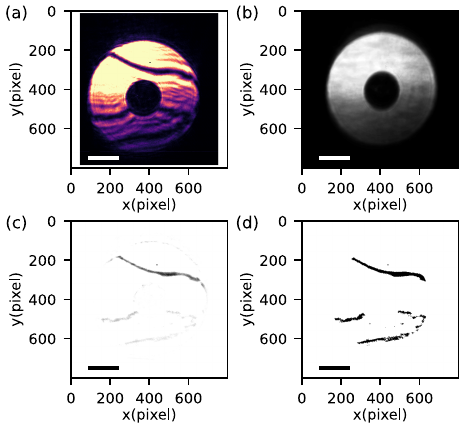}

\caption{Preprocessing steps. (a) Raw detector image. (b) Mask generated from 31 raw images. (c) Image after application of the mask, suppressing the background and nonfeature dark pixels. (d) Fully preprocessed image. The solid scale bars indicate 5~$\mu$m.}

\label{fig3}

\end{figure}

After recentering, a masking procedure is applied to suppress background intensity and interference patterns while preserving the magnetic domain walls. The domain walls appear as dark wavy lines that vary from image to image, and the optical background remains relatively constant within a scan line [see Fig.~\ref{fig3}(a)], so a mask can be generated by averaging all images within a given scan line.

Figure~\ref{fig3}(b) shows a representative mask obtained from the average of a scan-line dataset. Subtracting this mask from each individual image removes the background gradient and weak interference features while retaining the domain-wall contrast [Fig.~\ref{fig3}(c)].

\begin{figure}[t]

\includegraphics[width=1\columnwidth]{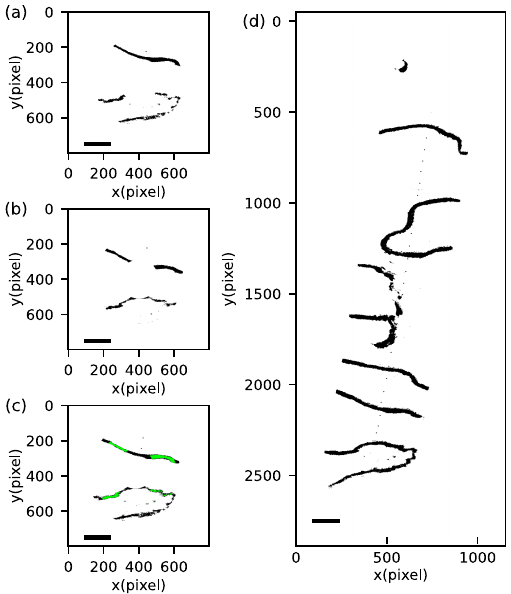}

\caption{Individual-image stitching. (a) and (b) Two adjacent preprocessed images. (c) Overlapped image with coincident domain-wall pixels highlighted in green. (d) Column image constructed from the preprocessed images. The solid scale bars indicate 5~$\mu$m.}

\label{fig4}

\end{figure}

\subsubsection{Trimming and binarization}

Following masking, residual circular boundaries remain because the mask and image are not perfectly aligned [see Fig.~\ref{fig3}(c)]. To remove these artifacts, each image is radially trimmed by excluding pixels inside a small inner radius and outside a larger outer radius of the donut-shaped profile.

The trimmed images are subsequently binarized by assigning pixels darker than a threshold value to black and all remaining pixels to white. This procedure eliminates intensity variations among domain walls and converts the images into uniformly weighted feature maps suitable for automated overlap analysis. A representative binarized image is shown in Fig.~\ref{fig3}(d).

\subsection{Stitching}

Following preprocessing, the images are stitched using a hierarchical reconstruction procedure. The stitching process is performed in two stages. Images acquired within the same raster scan line are first combined into column images, after which the resulting column images are stitched together to form the final large-area composite image. In both stages, image registration is achieved by maximizing the overlap of domain-wall features between neighboring images.

\begin{figure*}[t]

\includegraphics[width=\textwidth]{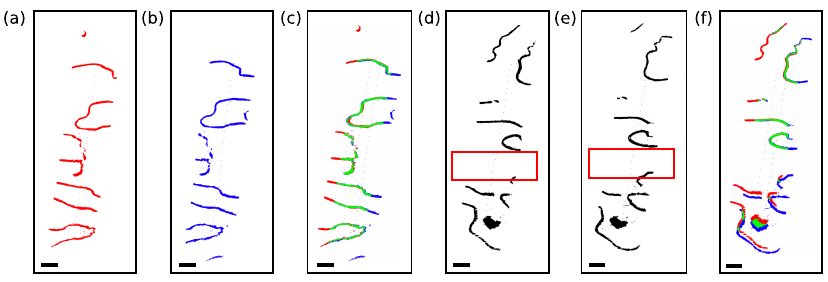}

\caption{Column-stitching procedure. (a) and (b) Two neighboring column images, color coded for clarity. (c) Composite image constructed from the two columns. (d)--(f) Representative example of a stitching error caused by insufficient overlapping features during the individual-image stitching stage. The solid scale bars indicate 5~$\mu$m.}

\label{fig5}

\end{figure*}

\subsubsection{Individual-image stitching}

The first stage of stitching combines images within a single scan line into a column image by determining the relative offsets between neighboring images. For each pair of adjacent images, one image is translated relative to the other over the allowed displacement range. At each trial displacement, an overlap score is calculated from the number of coincident domain-wall pixels. The displacement yielding the highest overlap score is selected as the optimal offset and recorded. The two images are then stitched together while preserving all identified domain-wall pixels.

Figures~\ref{fig4}(a) and \ref{fig4}(b) show two adjacent images within the same scan line, and Fig.~\ref{fig4}(c) shows the overlapped image with the overlapping domain-wall pixels corresponding to the highest overlap score highlighted in green. This procedure is repeated for every pair of adjacent images within the scan line to construct a single column image, as shown in Fig.~\ref{fig4}(d). Repeating the procedure for all 14 scan lines produces a total of 14 column images.

\subsubsection{Column stitching}

The second stage combines the 14 column images into a single large-area reconstruction. The same overlap-based algorithm is used to determine the relative offsets between neighboring column images. Adjacent columns are translated with respect to one another, and an overlap score is calculated for each trial displacement. The displacement yielding the highest overlap score is selected as the optimal offset and recorded. The aligned columns are then merged, and this procedure is repeated iteratively until a single composite image is obtained. Figures~\ref{fig5}(a) and \ref{fig5}(b) show two neighboring column images, and Fig.~\ref{fig5}(c) shows the resulting stitched image after alignment using the optimal offset.

During this process, we found that individual-image stitching occasionally produces an incorrect alignment when neighboring images contain too few common features. Such errors propagate to the corresponding column image and become apparent during the column-stitching stage. Figures~\ref{fig5}(d)--\ref{fig5}(f) illustrate a representative example. The highlighted region (red box) in Fig.~\ref{fig5}(e) contains sufficient common features for successful stitching, whereas the corresponding region in Fig.~\ref{fig5}(d) contains too few overlapping features. As a result, one image within the scan line for Fig.~\ref{fig5}(d) is incorrectly omitted during the individual-image stitching process, producing an erroneous column image. When the neighboring column images are subsequently stitched together, the side of the column with fewer features become misaligned as shown in the lower portions of the columns in Fig.~\ref{fig5}(f). This mismatch originates from the incorrect alignment introduced during the individual-image stitching stage due to the lack of sufficient overlapping features.

The fully automated column-stitching procedure may therefore generate stitched images containing errors, such as that shown in Fig.~\ref{fig5}(f). Such failures cannot be completely eliminated, so we manually omit the erroneous columns and repeat the column-stitching process. This approach successfully produces the final large-area reconstruction shown in Fig.~\ref{fig6}(a).

\subsection{Post-processing and final composite image}

\begin{figure}[t]

\includegraphics[width=1\columnwidth]{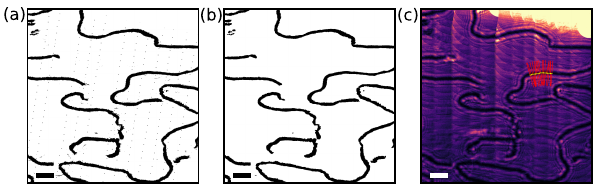}

\caption{(a) Final composite image constructed from the processed images and cropped to the region of interest. (b) Post-processed composite image after removal of residual detector artifacts. (c) Composite image reconstructed from the original images using the coordinates determined by the stitching procedure. Yellow dots indicate the positions at which the line cuts were centered, and red lines indicate the one-dimensional line cuts oriented locally perpendicular to the domain wall. The solid scale bars indicate 5~$\mu$m.}

\label{fig6}

\end{figure}

The final stage of the workflow removes residual detector artifacts. The dotted lines visible in Fig.~\ref{fig6}(a) originate from dead detector pixels present in individual images. The detector artifacts occur at the same detector coordinates, so they are repeated across all 31 images within a scan line, producing the dotted-line pattern in the stitched image. Such isolated dead pixels that survive the preprocessing and stitching procedures are eliminated using a morphological opening operation.\cite{8,9}

The opening transform first erodes image features using a kernel larger than the dead-pixel size, removing isolated detector artifacts. A subsequent dilation step\cite{8,9} restores the remaining domain-wall features while preserving the removal of the dead pixels. Because the detector artifacts are much smaller than the magnetic domain structures, this procedure has minimal effect on the reconstructed domain-wall network. The resulting post-processed image is shown in Fig.~\ref{fig6}(b).

Although the post-processed image in Fig.~\ref{fig6}(b) accurately reproduces the large-scale AFM domain-wall network, it does not retain the fine intensity variations, detailed domain-wall profiles, or interference fringe patterns present in the original measurements because the stitching algorithm operates on preprocessed images. To recover these details, we use the image coordinates determined during the stitching process to reconstruct the composite image directly from the original raw images. We then apply alpha blending between images to reduce intensity differences between the stitched columns. The resulting high-fidelity reconstruction is shown in Fig.~\ref{fig6}(c).

\begin{figure}[t]

\includegraphics[width=\columnwidth]{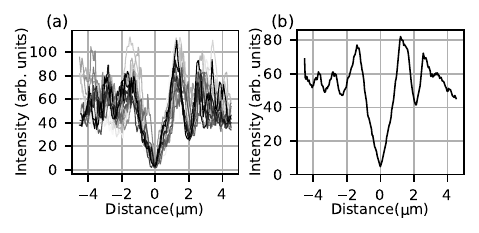}

\caption{(a) One-dimensional line cuts extracted at multiple locations marked in Fig.~\ref{fig6}(c). (b) Average of the line cuts shown in (a).}

\label{fig7}

\end{figure}

To verify that the original image quality of the experimental data is preserved while maintaining the accurate spatial alignment established by the automated stitching procedure, we examined one-dimensional line cuts across a domain wall, focusing particularly on a region where multiple images overlap, as shown in Fig.~\ref{fig6}(c). Line cuts were obtained perpendicular to the domain wall at multiple positions, as indicated in Fig.~\ref{fig6}(c) and shown in Fig.~\ref{fig7}(a). The average line profile is plotted in Fig.~\ref{fig7}(b). The resulting profile is consistent with previous measurements of domain walls in MnBi$_2$Te$_4$, exhibiting a deep valley corresponding to a domain wall, along with shallow oscillations characteristic of the interference pattern surrounding the domain wall.\cite{13,14}

\section{Summary}

We have developed MOSAICX, an automated workflow for stitching large direct-CXI datasets acquired using Fresnel zone plate optics. The method combines image centering, masking, trimming, binarization, hierarchical stitching, and post-processing to generate large-area composite images from raster-scanned measurements. Application of MOSAICX to 434 direct-CXI images of MnBi$_2$Te$_4$ successfully reconstructs the magnetic-domain and domain-wall landscape over an area substantially larger than the field of view of an individual image. By automating the alignment and reconstruction process, the workflow reduces the effort required for analyzing large direct-CXI datasets while minimizing stitching artifacts arising from image misalignment and detector defects. The approach is broadly applicable to raster-scanned direct-CXI measurements and provides a practical tool for visualization and quantitative analysis of mesoscale magnetic-domain structures.

\begin{acknowledgments}

This work was supported by the University of Wisconsin--Milwaukee. We gratefully acknowledge Claudio Mazzoli and Rahul Jangid for their invaluable assistance with the CXI measurements.

\end{acknowledgments}

\section*{Data Availability}

The data and code supporting this work are available at the GitHub repository described in Ref.~\cite{Starr}.

\bibliography{Stitching}

\end{document}